\documentstyle[prl,twocolumn,aps]{revtex}
\global\topskip -.3truein

\def\lm{\lower 7pt\hbox { $\lim$} \atop
{\eta\rarrow 0}}

\def\t0{\tau_{_0}}

\def\beq{\begin{equation}}
\def\eeq{\end{equation}}

\title{Low-Temperature Dephasing in Disordered Conductors:\\
the Effect of ``1/f'' Fluctuations}
\author{Yoseph Imry$^1$, Hidetoshi Fukuyama$^2$ and  Peter Schwab$^3$}
\address{$^1$Weizmann Institute of Science, Department of Condensed Matter
Physics,\\ IL-76100 Rehovot, Israel; Physics Department.
Faculty of Science, $^2$Tokyo University, Hongo,Bunkyo-ku, Tokyo 113 Japan;
$^3$Dipartimento di Fisica,  Universit\`a degli Studi di Roma ``La Sapienza'',
Piazzale Aldo Moro 2, I-00185 Roma, Italy. }
\date{\today}
\begin{document}

\maketitle
\begin{abstract}
Electronic quantum effects in disordered conductors are controlled by
the dephasing rate of conduction electrons. This rate
is expected to vanish with the temperature.
We consider the very intriguing recently
reported  apparent saturation of this dephasing rate in
several systems at very low temperatures. We show that  the ``standard model'' of
a conductor with static defects can {\em not} have such an effect. However, allowing
some dynamics of the defects may produce it.
\end{abstract}

\section{Introduction}
\label{Introduction}
\par
Electronic transport in real conducting materials at low temperatures
is now known \cite{book} to exhibit important quantum phenomena. Even in the
macroscopic limit, weak localization and the associated magnetoresistance \cite{LR}
are by now commonly observed. Sample-specific conductance fluctuations \cite{Alt}
are relevant  in mesoscopic samples at low temperatures.
The phase coherence of the electrons must be preserved over the relevant
length scales to maximize these effects. Thus, the precise understanding
how the phase coherence of the electrons is lost, is a crucial issue in mesoscopic
physics (as well as, obviously, in other branches of physics).
One of the important insights \cite{pc}
gained from these studies  is that elastic scattering {\em does
not} destroy phase coherence. It takes {\em inelastic scattering}
changing the quantum state of other degrees of freedom to do that.
Thus, the dephasing
(sometimes called "decoherence")
of Quantum-Mechanical interference occurs by
inelastic scattering
\footnote{Here
the term "inelastic" implies
just changing the quantum state of the environment. It is irrelevant how much
energy is transferred in this process.
This includes zero energy transfer -- flipping the environment
to a degenerate state, if that is possible.}
off the
degrees of freedom which do not directly participate and are not measured in the interference experiment \cite{Fey,SAI}.
Such degrees of freedom are often referred to as the "environment". This dephasing is
characterized by the ``phase-breaking" time, $\tau_\phi$, for the electron.
This phase-breaking time generically increases
\cite{AAK} with decreasing temperatures and diverges when $T \rightarrow 0$,
because of the decreasing phase space available for inelastic scattering.
This paper is devoted to the problem of the apparent saturation of  $\tau_\phi$
at low temperatures, recently observed in disordered metals\cite{webb,wisqm}.
We show \cite{CI} that  in the ``standard model'' of
a conductor with static defects, $\tau_\phi$  must in fact diverge when
$T \rightarrow 0$.
However, taking into account the relaxational
dynamics of the scattering centers may produce a finite, weakly temperature
dependent
$\tau_\phi$ at
very low temperatures, but normally not in the strict $T \rightarrow 0$
limit. It is the same defect dynamics which produces the
intimately related low-frequency (1/f) conductance noise.

\par
In the standard picture, one considers electrons performing
diffusive motion due to
defects, just
above the Fermi energy,
and interacting via the Coulomb interaction  with the other electrons.  It is
straightforward to obtain the dephasing rate
from the strength of the  inelastic scattering of the considered electron
by the environment, i.e.
using the trace left in the latter \cite{SAI}. A similar result
was first obtained in ref.\cite{AAK} by using the effect of the
electromagnetic fluctuations due to the electron gas on the considered
electron.
The equivalence \cite{SAI} of these two points of view is guaranteed by the
fluctuation-dissipation theorem.
\par
It is easy to manipulate these cited expressions in the semiclassical picture,
for a general $V_q$, into:
\begin{equation}
1/\tau_\phi = \int \int d{\bf q} d \omega |V_q|^2 S_p(q,\omega)
S_{env}(-q,-\omega).
\label{magic}
\end{equation}
where $S_p(q,\omega)$ is the dynamic structure factor of the diffusing
electron  (a Lorentzian with width $Dq^2$) and $S_{env}(-q,-\omega)$ is the
same for the environment. Evaluating these integrals
with the environment being the electron gas, yields the well-known
expressions found first by Altshuler et al. \cite{AAK}.
For 2D and 1D (thin films and wires)  some care is needed
to eliminate the infrared (small q)-divergence by considering  the phase
{\em difference} of two paths \cite{Fey,AAK}.
These results are in a {\em quantitative agreement}
with experiments, except for the low-temperature limit which we discuss
below. We remark that
the theory described here for dephasing is perturbative in the electron-environment
interaction. Since $\tau_\phi$ is much longer than the elastic scattering time,
$\tau$, the dephasing is weak and there is no need to invoke nonperturbative
ideas.
\section{Statement of the problem, proof that zero-point motion
does not dephase}
\label{II}
\par
Recently, Mohanty et al\cite{webb} have published extensive experimental data
indicating
that contrary to general theoretical expectations and to the results
mentioned above, the
dephasing rate in films and wires does not vanish as $T \rightarrow 0$.
Serious precautions\cite{wisqm} were taken to eliminate experimental
artifacts. It was speculated that
such a saturation of the dephasing rate when $T \rightarrow 0$, might
follow from interactions with the zero point motion of the environment. These
speculations have received apparent support from calculations in
ref.\cite{zaikin}. However, the latter were severely criticized in
refs.\cite{ale,CI,alt} and were in disagreement with experiments in
ref.\cite{kha}. In fact, it is clear that since dephasing must be
associated with a change of the quantum state of the environment, it
cannot happen as $T \rightarrow 0$ (a macroscopic degeneracy of the environment's
ground state is not considered here). In that limit neither the electron nor the
environment has any energy to exchange. This qualitative
argument can be made into a more formal proof\cite{CI,isqm}. While demonsrating
that zero point motion does not dephase, this proof does show what
{\em further} physical assumptions can in fact
produce a finite dephasing rate at very low temperatures.
\par
The proof\cite{CI} uses eq.\ref{magic} and applies the very general
detailed-balance relationship
\begin{equation}
S(q,\omega) = S(-q,-\omega) e^{-\hbar \omega /k_B T},
\label{db}
\end{equation}
to either   $S_p(q,\omega)$
or $S_{env}(-q,-\omega)$.
It is immediately seen that the integrand
of eq.\ref{magic} is a product of two factors one of which
vanishes for $\omega > 0$  and the other for $\omega < 0$, as $T\rightarrow 0$.
Thus the integral and the dephasing rate vanish in general when $T\rightarrow 0$. However, if, for example, $S_{env}(-q,-\omega)$ has an
approximate delta-function peak
at small $\omega$ due to an abundance of low-energy excitations,
one may get a finite dephasing rate at temperatures higher
than the width of that peak. Should $S_{env}(-q,-\omega)$ have an $1/\omega$
behavior at low $\omega$, the dephasing rate would have a $logT$
term at the corresponding temperatures. Such near-degeneracies of
the ground state are known to exist in disordered, glassy, systems. These
follow from the many ``mesoscopic'' realizations of the disorder configuration.
The system slowly fluctuates among these many states and it may in fact not
be in full equilibrium. This may cause \cite{1/f}  the  commonly observed
low-frequency (often ``$1/f$'') noise \cite{Weis,1/fe}. To include the
effects of such fluctuations on the dephasing, one  has, in principle,
just to add their relevant contribution to the full $S_{env}(-q,-\omega)$.
\par
An extremely  useful phenomenological  model for such rearrangements that
has been  employed to
explain the low-temperature properties of glasses \cite{Pohl}, considers\cite{AHV}
two-level tunneling systems (TLS). For a review, see refs.\cite{Osh1,yuri}.
\par
One may consider the dephasing due to the TLS
by either obtaining
the appropriate ineleastic scattering rate of the conduction electron off the TLS
or by adopting a time-domain picture. According to the latter, one
considers two interfering paths, and when enough motion of the TLS occurs
during the time to execute the paths, the interference is lost. As mentioned above, these two
descriptions are equivalent. We shall use the first for a simple calculation
but the second may be useful in the physical argumentation below. It is the
phase shift due to the motion of the scatterers which produces both the
conductance change and the dephasing. These occur, respectively, when the
defect motion is slower or faster than the time scales of the electronic
motion \cite{felix}.

\section{General "1/f" considerations}
\label{III}
\par
We start with a rather generic picture, strongly related  to 1/f
noise \cite{1/f}. We consider  those defects
that are rearranging, for example, by tunneling at low temperatures  at a rate
$\nu$
satisfying:
\beq
\hbar / \tau_{\phi} << h \nu << k_B T,
\label{range}
\eeq
where $\tau_{\phi}$ is the dephasing time.
The second inequality
reflects the fact demonstrated above, that transfers of energies $>> k_BT$
are cut off. The first inequality can be justified from an uncertainty principle
argument.
It is even more conveniently understood from the time-domain
point of view: as mentioned above, to decohere a given path, one needs
motions that occur on time
scales shorter than its duration. A discussion of the fact that fast fluctuations yield dephasing and slow ones yield
conductance fluctuations can be found in ref.\cite{felix}.
\par
We  denote the fraction of the defects that move on the time scales of
eq.\ref{range} by
$p <<1$. According to the standard theory for the $1/f$ noise, if $\nu$
is an exponential of (minus) a large dimensionless number which is
distributed more or less uniformly, the distribution of $\nu$ is
$1/(\nu \lambda)$ where $\lambda$ is a normalization constant.
The noise power per a
decade of frequency \cite{Weis} is therefore proportional to the
fraction of defects participating in the tunneling motion, $p_t$, divided
by $\lambda$. Therefore, $\lambda$ is given,
roughly, by the number of decades for which $1/f$ noise prevails.
Thus, for the relevant frequency range of eq.\ref{range},
\beq
p = \frac{p_t }{\lambda} log (\frac{\tau_{\phi} k_B T}  {\hbar}).
\label{p}
\eeq
Their density
is $n_0 = p n_i$ where $n_i$ is the total defect
density. The density of active centers per frequency decade, $n_1$
is smaller by the logarithmic factor, i.e.
$n_1 = n_i p_t/\lambda$
\par
We now briefly review the mesoscopic fluctuation model of Feng, Lee and Stone \cite{1/f}
for $1/f$ noise.
One  considers the system as the temperature is lowered
and the "standard-model" dephasing length, $L_{\phi}$, is increasing
\cite{Blanter}. We consider the 3D case, which is the worse for our purposes.
In refs. \cite{AS,1/f} it was found that the change of the dimensionless conductance of
a coherence volume due to moving a single defect with an effective  microscopic scattering
cross-section $\sigma$ a distance $d \alt 1/k_F$ is:
\beq
(\Delta_{1} g)^{2}= \frac{1}{(k_F \ell)^2} \frac{\ell}{L_{\phi}} F.
\label{change}
\eeq
The factor F is a product of two possible correction factors $F_1$ and $F_2$:
when the impurity moves a
distance $d<< (1/k_F)$, $F_1$ becomes the factor $(k_Fd)^2$ and if the impurity
scattering cross section $\sigma <<  (1/k_F)^2$, $F_2$
becomes $\sigma k_F^2$.
We also remember that under $1/f$ conditions, the density of defects in any
frequency window between $\nu$ and $\alpha \nu$ is the same, $n_1 log\alpha$,
independently of $\nu$, for a
constant  $\alpha$. For a given temperature, and  $L_{\phi}$, we choose $\alpha$
so that the change of the dimensionless conductance, $g$, for monoenrgetic electrons
due to all defects in the above frequency window and in the coherence volume
$L_{\phi}^d$ atains its full value of order unity. This determines $\alpha$ to be given
in 3D by:
\beq
1/ log\alpha =  (\frac{L_{\phi}}{\ell})^2 \frac{p_t}{\lambda} F_1.
\label{alpha}
\eeq
>From the above estimates and ref.\cite{1/f}, one finds for the
relative $1/f$ noise power
per decade in a macroscopic bulk sample with N electrons, by averaging the
noise over the many coherent volumes and over an energy band of $k_B T$:
\beq
\frac{(\Delta g)^2}{g^2} \sim \frac{L_{\phi}^3}{N\ell^3}
\frac{p_t}{\lambda k_F \ell} \frac{\hbar}{\tau_{\phi} k_B T} F_1,
\label{Hooge}
\eeq
where
the last factor,
valid when the ratio $\frac{\tau_{\phi} k_B T}{\hbar} \gg 1$,
is due to energy-averaging \cite{1/f}.
We shall see that typical values of the parameter $\frac{p_t}{\lambda}$
can be on the order of $10^{-3}$.
This implies that the Hooge parameter can reach the order of $10^{-2}$
in reasonably good metals ($k_F \ell \sim 100$) at low temperatures.
Much larger values of the Hooge parameter are possible for dirtier metals.
Interestingly, values of that parameter between one and ten have already been found
experimentally \cite{1/fe} for very dirty metals at low temperatures.
Should $L_{\phi}$  saturate at very low  temperatures, the temperature dependence of the
Hooge parameter would then convert from that of eq.\ref{Hooge} to a simple
proportionality to $1/T$. The latter should be valid as long as
$\frac{\tau_{\phi} k_B T}{\hbar} \gg 1$.
This might offer another way to observe the saturation of  $L_{\phi}$.
The dependence of the Hooge parameter on
$k_F \ell$ should also be modified upon saturation of the dephasing owing to losing
the  dependence of $L_{\phi}$ on purity.
\par
We now consider the frequency range of eq. \ref{range}.
Our principal physical observation is that when $L_{\phi}$ becomes large enough so that
the conductance fluctuation due to motions in all that range attains its full value of
order $e^2 / \hbar$, an important crossover occurs in the system.
This crossover temperature, $T_0$ is given, as in eq. \ref{alpha}, by
\beq
(\ell / L_{\phi})^2 \sim \frac{\tau}{\tau_\phi}
\sim \frac{p_t}{\lambda} F_1 log(\frac{\tau_{\phi} k_B T}{\hbar}).
\label{xover}
\eeq
Adopting the physical path-integral picture described in ref.\cite{1/f},
it is seen that below the crossover
temperature, eq. \ref{xover}, essentially all paths crossing a coherent
region (within the standard model) in the system, encounter and are
significantly phase-shifted by, defects moving at the relevant
rate. Therefore we expect that {\em eq.\ref{xover} should  give
the crossover to the regime where the
dephasing is {\em dominated by defect motion}. Below $T_0$, $1 / \tau_{\phi}$
is roughly saturated (apart from the logarithmic correction) and given by
its standard-model value at the crossover temperature, eq.\ref{xover}}.
This statement,
which is (see below) in a order-of-magnitude agreement with the results of ref.\cite{webb},
is consistent with but more general than, particular models
that can be considered
and will be treated below. It substantiates the belief that the defect dynamics is indeed
the explanation for the intriguing results of ref.\cite{webb}.
The above is valid as long as $\hbar / \tau_{\phi} < k_B T$. $1 / \tau_{\phi}$
should approach zero at temperatures below  the saturated $\hbar / \tau_{\phi}$.
\section{Model calculations}
\label{IV}
\par
We now perform a more detailed evaluation within the TLS model.
The Born approximation scattering cross section from a particle in a
double-minimum potential is easily
calculated, see for example ref.\cite{67}.
The tunnel splitting of the lowest two levels
in the double well is denoted by $\Omega$ and the asymmery
between the wells by $B$. For $B >> \Omega$, the various matrix elements squared
for transitions become  of the order of $(\Omega/B)^2$. Therefore, one may start with
the case of $ B\sim \Omega$, in which the qualitative behavior is that of a
symmetric well, which will be given below.
We shall later treat in more detail the important role of B.
The elastic and inelastic scattering
cross sections are given, up to the same constant, for momentum
transfer {\bf q}, by:
\beq
\sigma_{el}(q,\omega) \sim  cos^2(q.d)\delta(\omega),~
\sigma_{in}(q,\omega) \sim sin^2(q.d) \delta(\omega \pm \Omega).
\label{scatt}
\eeq
Here {\bf d} is the vector separating the two minima.
The $\pm$ signs in $\sigma_{inel}$ reflect energy gain and loss of the
tunneling energy by the scattered particle, corresponding to situations
where the tunneling particle was in the initial symmetric or antisymmetric
states.  For $B >> \Omega$,  $\sigma_{inel}$ acquires an additional factor
of $(\Omega/B)^2$, as mentioned above.

\par
We write the dephasing rate $\frac{1}{\tau_0}$ by the TLS
as the sum of the diffusive and the ballistic portions
of the integral over $q$ in eq.\ref{magic}. We
start with the diffusive ($q<<1/\ell$) contribution.
We perform the integration in eq.\ref{magic}, using equation \ref{scatt}
(times $n_0$)
for the environment and the usual diffusive Lorentzian for the electron.
We denote the defect scattering cross section by  $a^2$, where $a$ is
a microscopic length (of the order of or smaller than $1/k_F$, in metals).
We remember that
$1/\ell = n_i a^2$. The result is:
\beq
[\frac{1}{\tau_0}]_d \sim \frac {pF_1} {\tau} \frac{1}{(k_F \ell)^4}.
\label{satd}
\eeq
\par
The ballistic ($q >>1$) contribution to $1 / \tau_0$ is easily obtained
from the cross section, eq.\ref{scatt} and the density of the relevant
scatterers. For $k_F \ell >> 1$:
\beq
[\frac{1}{\tau_0}]_b \sim
\frac {pF_{1}} {\tau}.
\label{satb}
\eeq
Perhaps surprisingly, the ballistic contribution is dominant for a good
metal, $k_F \ell >> 1$. Obviously, once the rate $1/\tau_0$
obtained from the TLS becomes comparable to the standard model
$1 / \tau_{\phi}$, the former will dominate and the effective dephasing
rate will be $1/\tau_0$. It depends
very weakly (logarithmically, by eq. \ref{p}) on the temperature.
And at even lower temperatures, once it becomes smaller than the temperature, it should vanish as well.
The result of eq.\ref{satb} agrees with the more general eq.\ref{xover}.
\par
To  get a
low temperature dephasing rate of $10^{-5}/ \tau$\cite{webb}, we need
a value of $p \sim 10{^{-3}}$. Thus, taking very roughly
$ log(\frac{\tau_{\phi} k_B T}{\hbar})
/\lambda \cong 0.1$, we need that $p_t \sim 10^{-2}$
of the defects participate in the $1/f$ noise.
The impurity density, for impurities whose
scattering length is atomic, is of the order of $k_F^2  / \ell$.
This  typically
corresponds to total impurity concentrations  of the order of $10^{-2}$. Thus,
we have to assume
a $ \sim$ 100 ppm concentration of  low- energy
``two-level'' defects at the relevant range,
to get a low temperature dephasing rate  comparable to
that
of ref. \cite{webb}. This is not a very unlikely possiblity.
The test whether it really exists would be, as we shall see later, by checking
if appropriate levels of $1/f$ conductance noise exist in the
system. Confirming that at microwave-frquencies would be a more stringent test.
The cross-over of the $1/f$ noise behavior discussed above, and its
correlation with the saturation of $\tau_\phi$ would also test our picture.
\par
For completeness, we now treat the TLS model  in some more detail, taking into account
more seriously the double well asymmetry, $B$. This may give us some insight
on,
and a rough  order of magnitude estimate for, the parameter $p_t$ introduced
above.
We take the distribution of $B$
to be flat between zero and $B_0$, where $B_0$ can be expected to be on the
order of $10^{-2}$ eV, for atomic motion\footnote{Obviously, there are cases,
especially in semiconductors, where the defect ``motion'' is due to
electronic charge fluctuations. There, the energy scales are larger.}.
As above, we take the distribution of $\Omega$ to
be $1/\Omega$ starting from some very small $\Omega_{min}$ up to some
cutoff $\Omega_{max}$. We shall see that for  $\Omega_{max}\ll k_BT$,
the dephasing rate is constant but it is easy to see that  it vanishes linearly with temperature once $\Omega_{max}\gg k_BT$.
We have to integrate the dominant ballistic
contribution as in eq.\ref{satd} over the above distributions of $B$ and
$\Omega$, $P(B,\Omega) = 1/(\Omega B_0 \lambda)$.
For $\hbar/\tau{_\phi} < \Omega_{max} \ll k_BT$, the contributions
of $\Omega < B$ and $\Omega > B$ are comparable and the result is:
\beq
\frac{\tau}{\tau_0} \sim \frac{\Omega_{max}}{B_0 \lambda} F_1.
\label{result}
\eeq
Thus the ratio $\frac{\Omega_{max}}{B_0}$ basically replaces (apart from
the logarithmic correction) the unknown
parameter $p_t$ above.
The very reasonable estimate  $\frac{\Omega_{max}}{B_0} \sim 10^{-4}$,
yields an order of magnitude in agreement with the results of ref. \cite{webb}.
For the different regime, $\Omega_{max} < \hbar/\tau_{\phi} \ll k_BT$, we find:
\beq
\frac{1 }{\tau_0} \sim
{\Omega_{max}}{(F_1 B_0 \hbar \lambda \tau )^{-1/2}},
\label{result'}
\eeq
which has an encouraging order of magnitude as well.
\section{Concluding remarks}
\label{V}
\par
Checking experimentally the value of the Hooge parameter
as a function of purity and temperature \cite{Cohen}
would be very valuable.
The crucial issue is
whether our mechanism is consistent with  the observed level of low-temperature ($1/f$) noise. The nontrivial requirement is that the roughly sufficient levels of such noise that occur at low frequencies
can persist up to
gigahertz frequencies. While this is not the standard
regime for $1/f$ noise, it must be remembered that {\em five orders of magnitude
in frequency may correspond to a factor of two in barrier height.}
Experimental studies of the possible
correlation between the saturation of $\tau_{\phi}$ and  the $1/f$ noise
(or associated two-level absorption) at $10^{8-9}Hz$, are crucial.
\par
We conclude that while the "standard model" of
disordered metals (in which the defects are strictly frozen)
gives of course an infinite $\tau_\phi$ as $T \rightarrow 0$, there may be other physical ingredients that can make $\tau_\phi$
finite at very low temperatures, {\em without}
contradicting any basic law of physics. The TLS model
used here is a  particular
example and its requirements may or may not be satisfied in the real samples.
But, other models with similar dynamics might exist as well, as suggested by the qualitative argument following eq.\ref{xover}.
We reemphasize that $\tau_\phi$ will in fact diverge in the
$T \ll \hbar / \tau_\phi$ limit, which is in fact a strong prediction that
can be checked experimentally.
Thus, our results do {\em not} imply dephasing by zero-point
fluctuations, which has been repeatedly, and wrongly,
claimed in the literature. The failure of the semiclassical approximation used
in those considerations was clarified in ref.\cite{CI}.
\par It should perhaps be emphasized that in a typical measurement, the fluctuations in
the whole frequency range between $1 / \tau_{\phi}$ and the inverse of the (macroscopic)
measurement time are averaged upon.
\par
It may be asked why the results reported in refs.\cite{ale,alt} for dirtier samples
do not exhibit saturation of $\tau_{\phi}$. One may speculate that this can
be due not only to the fact that the diffusive regime scattering from the TLS
is weaker, but, perhaps more importantly, to a possible crossover to a new regime
where correlations among the relaxation of different defects become important.
Such a crossover was recently invoked to explain the experimental results of
ref.\cite{Osh}. This should push the relaxation modes to lower frequencies
and thus make them less effective for dephasing.
\par
It may be asked whether the picture we have introduced is  necessarily
a nonequilibrium one. Clearly, on any time scale shorter than
$1/\omega_{min}$,
the (unknown) lower cutoff of the 1/f noise, the system is not in full
equilibrium. However, the modes having the relevant time-scales, of
eq.\ref{range}, may well be in a partial equilibrium. At these frequencies
the $S(q, \omega)$ of the system contains the contribution of these modes
which do not appear in (and are actually a small correction to) the Drude
contribution. This full $S(q, \omega)$, from eq.\ref{magic} is the {\em only}
input necessary to determine
$\tau_\phi$. We think that its portion at the relevant frequencies may
well be the equilibrium one. Obviously, such a contribution will show up
in saturable electromagnetic and ultrasonic absorption in the metal.
This will become easier to observe for smaller $k_F \ell$.
\par
We further mention that an apparent saturation of $L_{\phi}$ at low temperatures implies that the
{\em length-scale for quantum behavior remains finite when $T \rightarrow 0$}.
Thus, "quantum critical behavior" will be blocked at very low temperatures.
One may speculate that this can explain several recently observed low temperature phenomena: the metallic 2D behavior and
the effective rounding of the Quantum-Hall-to-insulator
and the superconductor-to-insulator transitions.

\section{ Acknowledgements}
Research at WIS was supported by grants from the German-Israel
Foundation (GIF) and the Israel Science Foundation, Jerusalem.
Some of this work was performed while YI was visiting ICTP Trieste,
CEA Orme des Merisiers and ITP Santa Barbara (where it was supported
by the NSF under grant no. PHY94-07194). HF would like to acknowledge
Grant-in- Aid from Monbusho and Hf and YI thank
the International Symposium on the Foundations of Quantum Mechanics
sponsored by Basic Research Lab of Hitachi, where discussions took place.
PS was supported by the TMR  program of the EU.
We thank Y. Aharonov, C.
Castellani, D. Cohen,  R. Raimondi and A. Stern for collaborations on related
problems. B.Altshuler, I. Aleiner,
N. Argaman, M. Berry, M. Devoret, D. Esteve, Y. Gefen, S. Girvin,
S. Girvin, C. Glattli,
A. Kapitulnik, Y. Levinson, D. D. Osheroff, Z. Ovadyahu, J.-L Pichard,
T.D. Schultz, A. Stern, R. A. Webb, H.A. Weidenm\"uller, P. W\"olfle and
A. Zaikin are thanked for discussions.

\end{document}